\documentclass[oldversion]{aa} 
\usepackage{graphicx}
\usepackage{txfonts,color, natbib}
\begin{document}
\newcommand{\mps}{m\,s$^{-1}$}
   \title{Large-scale horizontal flows in the solar photosphere}

   \subtitle{II. Long-term behaviour and magnetic activity response}

   \author{M. \v{S}vanda
          \inst{1,2}
          \and
          M. Klva\v{n}a
          \inst{1}
          \and
          M. Sobotka\inst{1}
          \and
          V. Bumba\inst{1}
          }

   \offprints{M. \v{S}vanda}

   \institute{Astronomical Institute, Academy of Sciences of the Czech Republic (v.~v.~i.), Fri\v{c}ova 298,
              CZ-251 65, Ond\v{r}ejov, Czech Republic\\
              \email{svanda@asu.cas.cz, mklvana@asu.cas.cz, msobotka@asu.cas.cz, bumba@asu.cas.cz}
         \and
             Astronomical Institute, Charles University in Prague, V Hole\v{s}ovi\v{c}k\'{a}ch 2,
             CZ-180 00, Prague~8, Czech Republic\\
             }

   \date{Received ; accepted }
  \abstract{Recently, we have developed a method useful for mapping large-scale horizontal velocity fields in the
  solar photosphere. The method was developed, tuned and calibrated using the synthetic data. Now, we applied the method
  to the series of Michelson Doppler Imager (MDI) dopplergrams covering almost one solar cycle in order to get the information about the long-term behaviour of surface flows. We have found that our method clearly reproduces the widely accepted properties of mean flow field components, such as torsional oscillations and a pattern of meridional circulation. We also performed a periodic analysis, however due to the data series length and large gaps we did not detect any significant periods. The relation between the magnetic activity influencing the mean zonal motion is studied. We found an evidence that the emergence of compact magnetic regions locally accelerates the rotation of supergranular pattern in their vicinity and that the presence of magnetic fields generally decelerates the rotation in the equatorial region. Our results show that active regions in the equatorial region emerge exhibiting a constant velocity (faster by $60 \pm 9$~\mps{} than Carrington rate) suggesting that they emerge from the base of the surface radial shear at $0.95\ R_\odot$, disconnect from their magnetic roots, and slow down during their evolution.

   }
\titlerunning{Large-scale horizontal flows in the solar photosphere II: Long-term behaviour\dots}

   \maketitle
%
\section{Introduction}
The largest scale velocity fields in the solar photosphere consist of rotation profile and a meridional flow pattern. 

Basically, the differential rotation is described as an integral of the zonal component $v_\varphi$ of the studied flow field. The integrated flow field may be obtained using spectroscopic method, using tracer-type measurements, or using helioseismic inversions. The latest case allows to measure the solar rotation not only as a function of heliographic latitude, but also as a function of depth. From the helioseismic inversion we know that throughout the convective envelope, the rotation rate decreases monotonically toward the poles by about 30~\%. Angular velocity contours at mid-latitudes are nearly radial. Near the surface at the top of the convection zone there is a layer of a large radial shear in the angular velocity. At low and mid-latitudes there is an increase in the rotation rate immediately below the photosphere which persists down to $r \sim 0.95~R_\odot$. The angular velocity variation across this layer is roughly 3~\% of the mean rotation rate and according to the helioseismic analysis of \cite{2002SoPh..205..211C} the angular velocity $\omega$ decreases within this layer approximately as $r^{-1}$, where $r$ is a radial coordinate. At higher latitudes, the situation is less clear. For the overview of solar differential rotation measurements see \cite{1985SoPh..100..141S} or a more recent review by \cite{2000SoPh..191...47B}. 

The \emph{torsional oscillations}, in which narrow bands of faster than average rotation, interpreted as zonal flows, migrate towards the solar equator during the sunspot cycle, were discovered by \cite{1980ApJ...239L..33H}. Later research \citep{2001ApJ...559L..67A} found that there exist two different branches of torsional oscillations. At latitudes below about 40\,$^\circ$, the bands propagate equatorward, but at higher latitudes they propagate poleward. The low-latitude bands are about 15\,$^\circ$ wide in latitude. The flows were studied in surface Doppler measurements \citep{2001ApJ...560..466U}, and also using local helioseismology \citep{1997ApJ...482L.207K}. The surface pattern of torsional oscillations penetrate deeply in the convection zone, possibly to its base, as suggested by \cite{2002Sci...296..101V}. The amplitude of the angular velocity variation is about 2--5~nHz, which is roughly 1~\% of the mean rotational rate (5--10~\mps). The direct comparison between different techniques inferring the surface zonal flow pattern \citep{2006SoPh..235....1H} showed that the results are sufficiently coherent.
 
The surface magnetic activity corresponds well with the torsional oscillation pattern -- the magnetic activity belt tends to lie on the poleward side of the faster-rotating low-latitude bands. The magnetic activity migrates towards the equator with the low-latitude bands of the torsional oscillations as the sunspot cycle progresses \citep{2004ApJ...603..776Z}. Some studies \citep[e.\,g.][]{2002ApJ...575L..47B} suggest that meridional flows may diverge out from the activity belts, with the equatorward and poleward flows well correlated with the faster and slower bands of torsional oscillations. In a recent theoretical study by \cite{2007ApJ...655..651R} it was suggested that the poleward-propagating high-latitude branch of the torsional oscillations can be explained as a response of the coupled differential rotation/meridional flow system to periodic forcing in midlatitudes of either mechanical (Lorentz force) or thermal nature. The equatorward-propagating low-latitude branch is most likely not a consequence of the mechanical forcing alone, but rather of thermal origin.
  
The axisymmetric flow in the meridional plane is generally known as the \emph{meridional circulation}. The meridional circulation in the solar envelope is much weaker than the differential rotation, making it relatively difficult to measure. 
 
Two principal methods are widely used to measure the meridional flow: feature tracking and direct Doppler measurement. There are several difficulties complicating the measurements of the meridional flow using tracers. Sunspots and filaments do not provide sufficient temporal and spatial resolution for such studies. Sunspots also cover just low latitudinal belts and do not provide any informations about the flow in higher latitudes. Doppler measurements do not suffer from the problem associated with the tracer-type measurements, however they introduce another type of noisy phenomena. It is difficult to separate the meridional flow signal from the variation of the Doppler velocity from the disc centre to the limb. Using different techniques, the parameters of the meridional flow show large discrepancies. It is generally assumed that the solar meridional flow in the close subphotospherical layers is directed poleward with one cell per hemisphere. Such flow is also produced by early global hydrodynamical simulation such as \cite{1982ApJ...256..316G}. As reviewed by \cite{1996ApJ...460.1027H}, the surface or near sub-surface velocities of the meridional flow are generally in the range 1--100~\mps, the most often measured values lie within the range of 10--20~\mps. The flow has often a complex latitudinal structure with both poleward and equatorward flows, multiple cells, and large asymmetries with respect to the equator. \cite{2004ApJ...603..776Z} used the time-distance helioseismology to infer the properties of the meridional flow in years 1996--2002. They found the meridional flows of an order of 20~\mps, which remained poleward during the whole period of observations. In addition to the poleward
meridional flows observed at the solar minimum, extra meridional circulation cells of flows converging toward the
activity belts are found in both hemispheres, which may imply plasma downdrafts in the activity belts. These
converging flow cells migrate toward the solar equator together with the activity belts as the solar cycle evolves. \cite{2002ApJ...575L..47B} measured the meridional flow (and torsional oscillations) using the time-distance helioseismology and found the residual meridional flow showing divergent flow patterns around the solar activity belts below a depth of 18~Mm. 

The most complete maps of the torsional oscillations and the meridional flow available at present have been constructed on the basis of Mt.~Wilson daily magnetograms \citep[see][]{ulrich90}. The measurements cover more than 20 years (since 1986) and the results obtained using this very homogenous material agree well with the properties described above.

The modern dynamo flux-transport models use the meridional flow and the differential rotation as the observational input. In the models by Dikpati et al. (\citeauthor{2006ApJ...638..564D} \citeyear{2006ApJ...638..564D} or \citeauthor{2006ApJ...649..498D} \citeyear{2006ApJ...649..498D}) the return meridional flow at the base of the convection zone is calculated from the continuity equation. They found the turnover time of the single meridional cell of 17--21~years. The meridional flow is assumed to be essential for the dynamo action, global magnetic field reversal and forecast of the future solar cycles.

There are known many relations of the differential rotation profile to the phase of the progressing solar cycle -- see e.\,g. \cite{2003SoPh..212...23J} or \cite{2005ApJ...626..579J} -- showing for example different properties of the differential rotation profile in the odd and even solar cycles. The rotation of the sunspots in relation to their morphological type was studied e.\,g. by \cite{1986AA...155...87B} who found that more evolved types of sunspots (E, F, G and H type) rotate slower than less evolved types. \cite{2004SoPh..221..225R} investigated Greenwich Photoheliographic Results for the years 1874--1976 and found a clear evidence for the deceleration of the sunspots in the photosphere with their evolution. \cite{2002aprm.conf..427H} found that the leading part of a complex sunspot group rotate about 3~\% faster than the following part. The dependence of the rotation of sunspot on their size and position in the bipolar region was investigated by \cite{1994SoPh..151..213D}. They explained the observed behaviour through a subtle interplay between the forces of magnetic buoyancy and drag, coupled with the role of the Coriolis force acting on rising flux tubes. This dynamics of rising flux tubes also explains the faster rotation of smaller sunspots. In average, sunspots rotate about 5~\% faster than the surrounding plasma.

In the theoretical study \citep{2004SoPh..220..333B} based on 3-D numerical simulations of compressible convection under the influence of rotation and magnetic fields in spherical shells, the author stated that in the presence of magnetic field the Maxwell stresses may oppose the Reynolds stresses and therefore the angular momentum is propagated more to the poles than without the presence of magnetic fields. As a consequence the rotation profile is more differential in the periods of lowered magnetic activity and it leads to the increase of the rotation in low latitudes. This behaviour was observed in many studies, e.\,g. \cite{1990ApJ...357..271H}.

The subject of this work is a verification of the performance of the method described in \cite{svanda06} (hereafter Paper~I) on the real data and the investigation of long-term properties of the flows at largest scales obtained with this method. We shall also discuss the influence of magnetic fields on the measured zonal flow in the equatorial region. This topic will be studied more in detail in one of the next papers in the series.

\begin{figure*}
\centering
\resizebox{\textwidth}{!}{\includegraphics{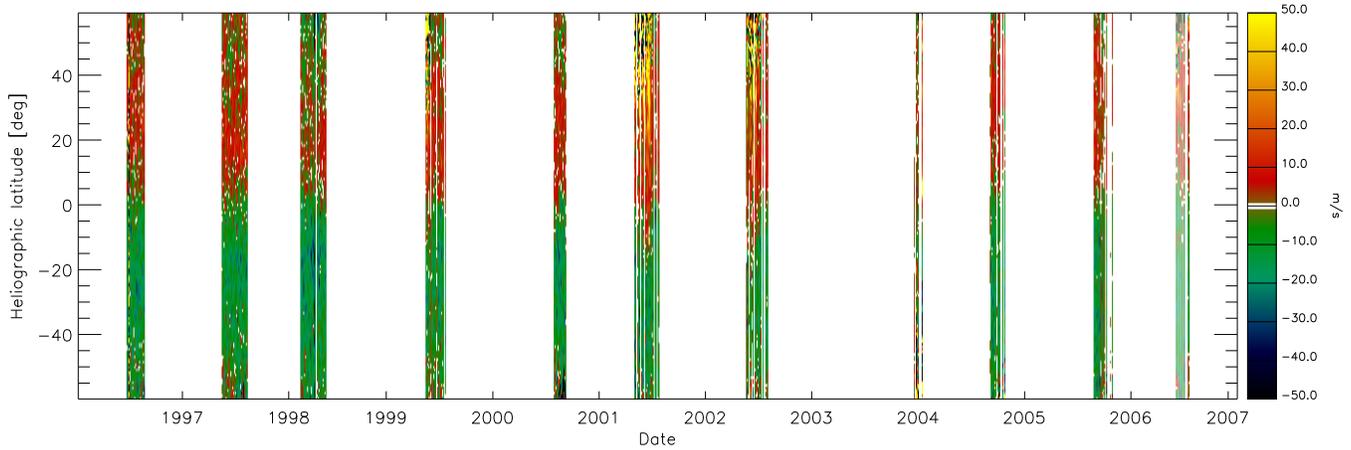}}
\caption{Mean meridional flow in time and heliographic latitude. It can be clearly seen that for almost all the processed measurements a simple model of one meridional cell per hemisphere would be sufficient. However, some local corruptions of this simple idea can be noticed on both hemispheres.}
\label{fig:meridional}
\end{figure*}

\begin{figure*}
\centering
\resizebox{\textwidth}{!}{\includegraphics{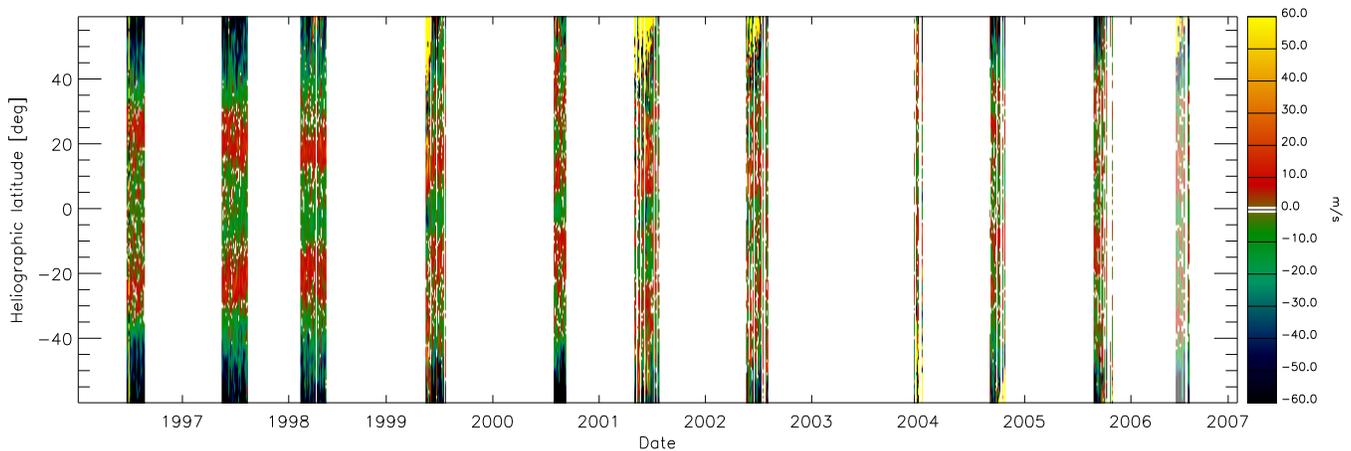}}
\caption{Torsional oscillations. The residua of the mean zonal flow with respect to its parabolic fit displayed in time and heliographic latitude. In the period of weak magnetic activity the pattern of belts propagating towards the equator is very clear. In the periods of stronger magnetic activity the flow field is influenced by the local motions in active regions and therefore the pattern of torsional oscillations is not clearly seen.}
\label{fig:torsional}
\end{figure*}

\section{Data processing}
The horizontal photospheric velocity fields are calculated using local correlation tracking (LCT) algorithm \citep{1986ApOpt..25..392N} applied to series of processed full-disc dopplergrams measured by Michelson Doppler Imager \citep[MDI; ][]{1995SoPh..162..129S} on-board Solar and Heliospheric Observatory (SOHO). In the dopplergrams, the supergranular pattern is tracked in order to obtain properties of the velocity field of the larger scale. 

For this study we have processed all suitable data measured by MDI. The instrument observed full-disc dopplergrams approximately two months each year in a high cadence of one frame per minute. These \emph{Dynamic campaigns} provide suitable material for our method. Between May 23, 1996 and May 22, 2006 we have 806 days covered by high-cadence measurements. In each of these days, two 24-hour averages sampled every 12 hours were calculated. In some days, MDI had significant gaps in measurements, so in such cases we did not have enough homogeneous material to process. Therefore, in all the \emph{Dynamic campaigns} 502 days were useful for our analysis and we calculated 1004 full-disc horizontal velocity fields. The processing of almost 3~TB of primary data took several months using fast computers running in the network of W.~W.~Hansen Laboratory, Stanford university. 

The data were processed using the technique described in detail in Paper~I. We bring just a brief summary. The method processes 24-hours series of MDI full-disc dopplergrams, containing 1\,440 frames. The one-day series first undergoes the noise and disturbing effects removal. From all frames, the line-of-sight component of the Carrington rotation is subtracted and the effect of a perspective is corrected. The frames are transformed so that the heliographic latitude of the disc centre $b_0=0$ and the position angle of the solar rotation axis $P=0$. Then, the $p$-modes of the solar oscillations are removed using a weighted average \citep[see][]{1988SoPh..117....1H}. The weights have a Gaussian form given by the formula:
\begin{equation}
w(\Delta t)=e^{\frac{(\Delta t)^2}{2a^2}}-e^{\frac{b^2}{2a^2}}\left(1+\frac{b^2-(\Delta t)^2}{2a^2} \right),
\end{equation}
where $\Delta t$ is a time distance of a given frame from the central one (in minutes), $b=16$~minutes and $a=8$~minutes. We sample averaged images in the interval of 15 minutes. The filter suppresses more than five hundred times the solar oscillations in the 2--4 mHz frequency band.

The processing of averaged frames consists of two main steps. In the first main step the mean zonal velocities are calculated and, on the basis of expansion to the Fay's formula $\omega=c_0+c_1 \sin^2 b + c_2\sin^4 b$, the differential rotation is removed. In the second main  step, the LCT algorithm with an enhanced sensitivity is applied. Finally, the differential rotation (obtained in the first step) is added to the vector velocity field obtained in the second main step. Both main steps can be divided into several sub-steps, which are mostly common.

\begin{enumerate}
\item The data series containing 96 averaged frames is ``derotated'' using the Carrington rotation rate in the first step and using the calculated differential rotation in the second step.

\item Derotated data are transformed into the Sanson-Flamsteed coordinate system to remove the geometrical distortion caused by the projection to the disc. The Sanson-Flamsteed (also known as sinusoidal) pseudo-cylindrical projection conserves the areas and therefore is suitable for the preparation
of the data used by LCT.

\item Remapped data undergoes the $k$-$\omega$ filtering \citep[e.~g.][]{1989ApJ...336..475T} with the cut-off velocity 1\,500~\mps{} for suppression of the noise coming from the evolutionary changes of supergranules, of the numerical noise, and for the partial removal of the ``blind spot'' (an effect at the centre of the disc, where the supergranular structures are almost invisible in dopplergrams due to the prevailing horizontality of their internal velocity field).

\item Finally,  the LCT is applied: the lag between correlated frames is 4~hours, the correlation window has \emph{FWHM} 60\arcsec, the measure of correlation is the sum of absolute differences and the nine-point method for calculation of the subpixel value of displacement is used. The calculated velocity field is averaged over the period of one day.

\item The resulting velocity field is corrected using the formula
\begin{equation}
v_{\rm cor}=1.13\,v_{\rm calc},
\end{equation}
where $v_{\rm calc}$ is the magnitude of velocities coming from the LCT, and $v_{\rm cor}$ the corrected magnitude. The directions of the vectors before and after the correction are the same.  The calibration formula was obtained from the tests on the synthetic data (see Paper~I). Finally, $v_x$ component is corrected for the data-processing bias of $-15$~\mps{} determined in Paper~I too. 
\end{enumerate}

\noindent In this study, we were interested only in the properties of the mean zonal and meridional components. Therefore, from each two-component horizontal velocity field the mean zonal and meridional component depending only on heliographic latitude were calculated as the longitudinal average of the flow map, using 135 longitudinal degrees around the central meridian. As stated in Paper~I, the  accuracy for each velocity vector is 15~\mps{} for velocities under 100~\mps{} and 25~\mps{} for velocities above 100~\mps. These inaccuracies have a character of a random error, therefore for the mean zonal and meridional components the accuracy is in the worst case 1~\mps.

The performance of the method was verified by \cite{2007astro.ph..1717S}. The results of comparisons between the technique used in the present study and the time-distance helioseismology show that both methods reasonably match. The calculated surface flows may be biased by the projection effects, although the tests on the synthetic data (Paper~I) did not show any signs of them. \cite{2006ApJ...644..598H} showed that the apparent superrotation of the structures tracked in dopplergrams reported by many studies can be explained as the projection effect. However, this bias would produce the systematic error, which should influence neither the periodic analysis, nor the relative motions of the active regions with respect to its surroundings.

\section{Results}

\subsection{Long-term properties}
For the study of long-term evolution of surface flows maps containing the mean zonal and meridional components were calculated. The maps of mean meridional component in time and heliographic latitude are shown in Fig.~\ref{fig:meridional}. We can clearly see that on the northern hemisphere dominates the flow towards the northern pole while on the southern hemisphere the flow towards southern pole prevails. The ``zero line'', the boundary between the flow polarity is not located exactly on the solar equator and seems to be shifted to the south in the period of increased solar activity (2001 and 2002). In agreement with \cite{1997AA...319..683M} and \cite{1998SoPh..183..263C} we found the meridional flow stronger in the periods of increased solar activity by about $~10$~\mps{} than in periods with lower magnetic activity.

A similar map was made in the same way for the zonal component. The mean equatorial zonal velocity for all the data is 1900~\mps. For all the processed data the dependence on latitude is close to a parabolic shape, parameters of which change slowly in time. The residua of the zonal velocity with respect to its parabolic fit given by
\begin{equation}
v_b=a_0+a_1 b + a_2 b^2,
\end{equation}
where $b$ is the heliographic latitude and $v_b$ the mean zonal velocity in the given latitude, were calculated in
order to see if we are able detect torsional oscillations in our measurements. As it is displayed in Fig.~\ref{fig:torsional}, the method clearly reveals torsional oscillations as an excess of the mean zonal velocity with respect to the zonal velocity in the neighbourhood. The behaviour of torsional oscillations is in agreement with their usual description -- the excess in magnitude is in order of 10~\mps, they start at the beginning of the solar cycle in high latitudes and propagate towards the equator with the progress of the 11-year cycle. However, with our method the visibility of torsional oscillations decrease with increasing solar activity. In the periods of strong activity both belts are not so clearly visible since the large-scale velocity field and its parabolic fit are strongly influenced by the presence of magnetic regions. However, the torsional oscillations belts still remain visible when the mean zonal component is symmetrised with respect to the solar equator. We did not focus on study of meridional flow or torsional oscillations depending on time and latitude, we just used them to check the ability and performance of our method.

\subsection{Periods in the mean components}
The mean zonal and meridional components in the equatorial area (averaged in the belt $b=-5\,^\circ - +5\,^\circ$) were analysed in order to examine the periods contained in the data. Since the data are not equidistant at all, we cannot use a simple harmonic analysis, so that the \emph{Stellingwerf method} \citep{1978ApJ...224..953S} was applied. It works on the principle of the phase dispersion minimization. The method sorts the data for every searched period into the phase diagram. Then the phase diagram is divided in a few (mostly ten) parts and for every part the mean dispersion is calculated. If the studied period is reasonable, the data-points group along the periodic curve and the dispersion in each part of the phase diagram is smaller than the dispersion of the whole data series. The normalized parameter $\theta \in \left(0,1\right>$ describes the quality of a given period. 

\begin{figure*}
\centering
\resizebox{0.49\textwidth}{!}{\includegraphics{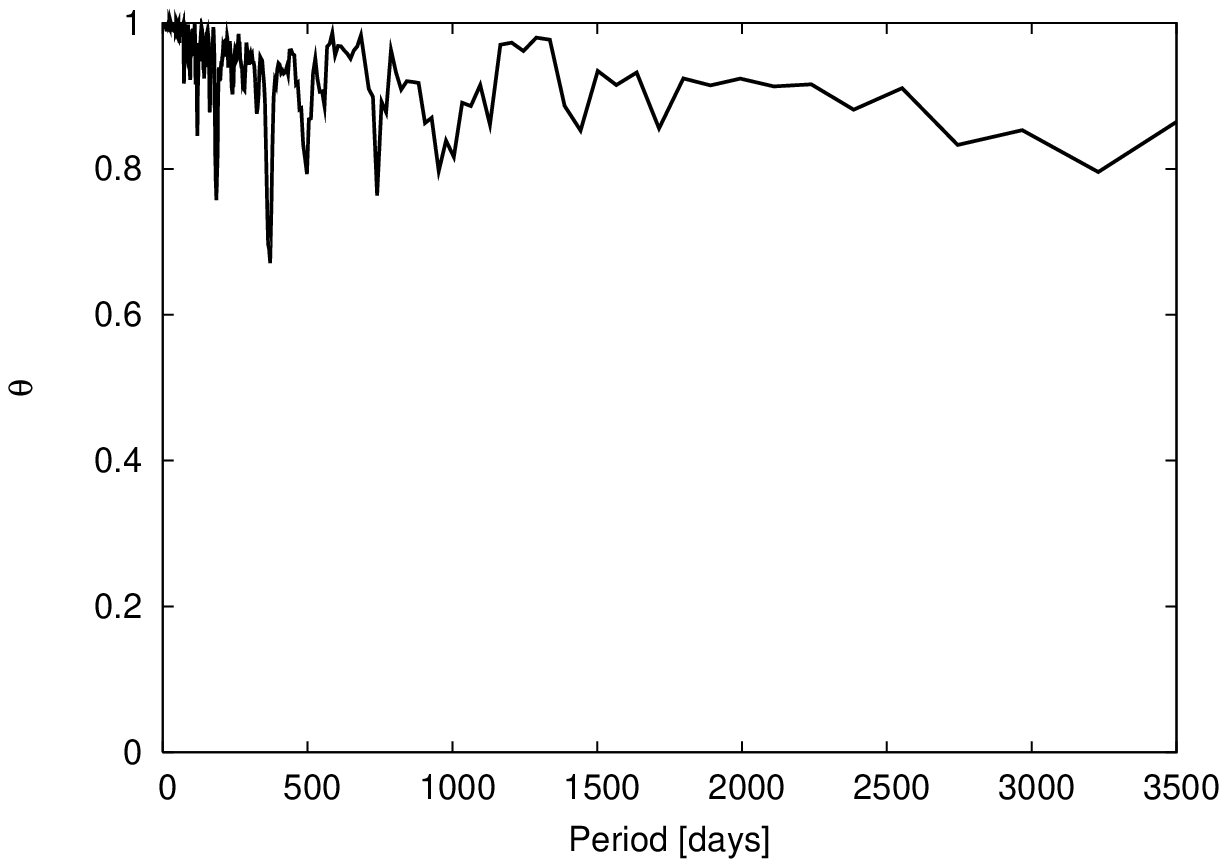}}
\resizebox{0.49\textwidth}{!}{\includegraphics{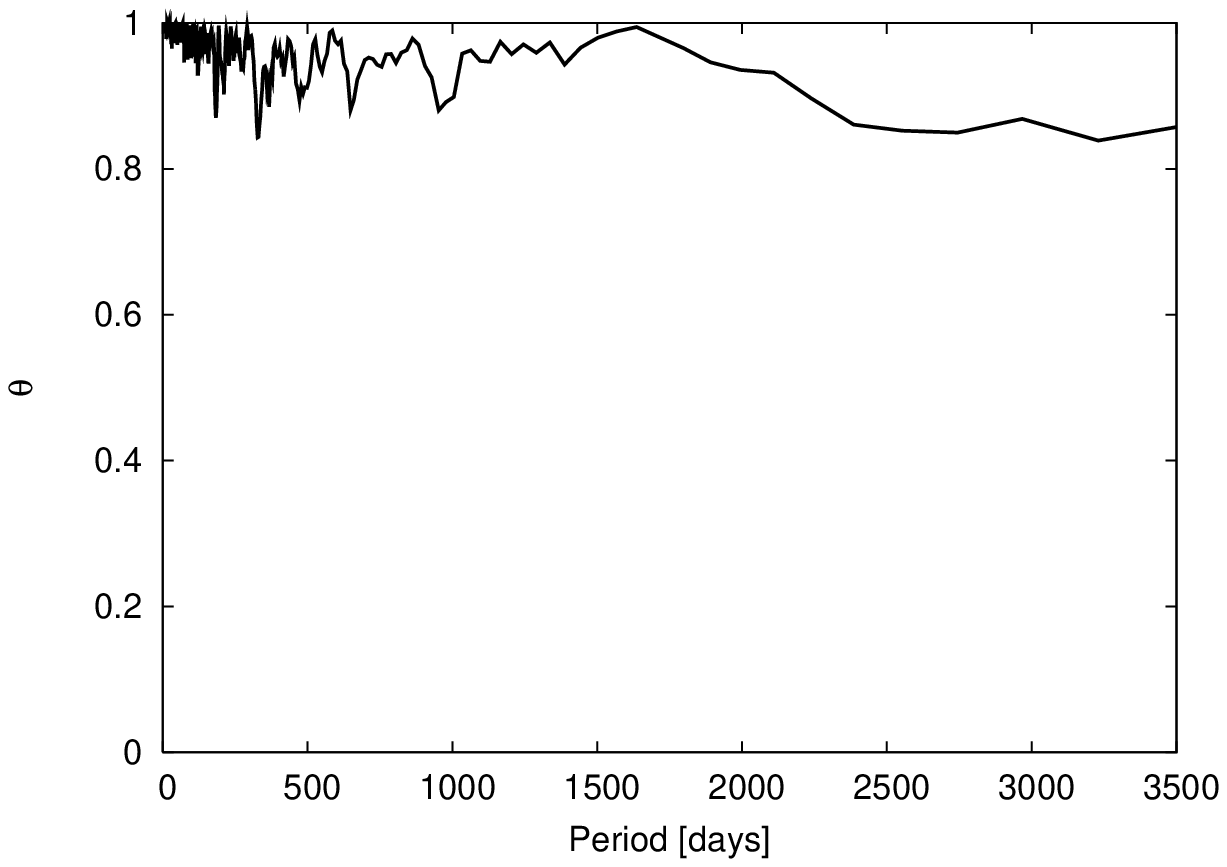}}\\
\resizebox{0.49\textwidth}{!}{\includegraphics{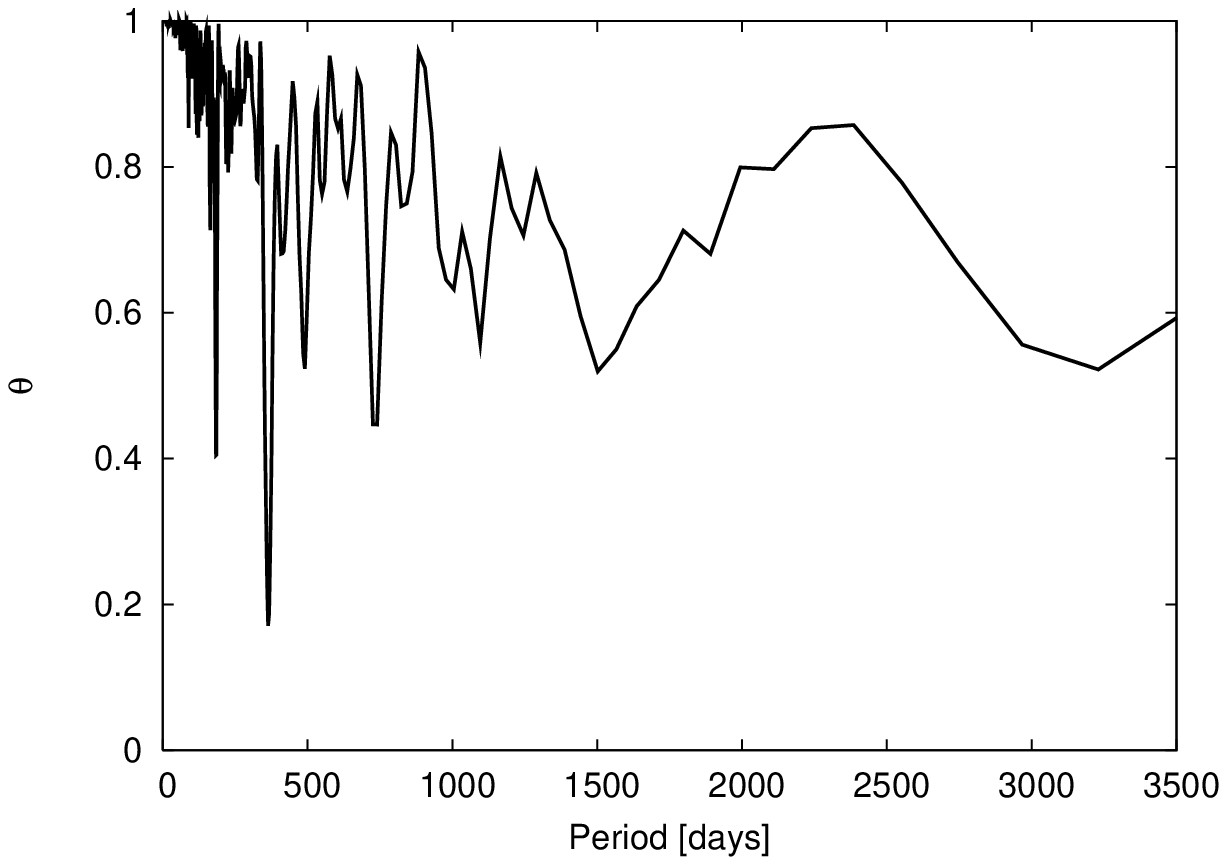}}
\resizebox{0.49\textwidth}{!}{\includegraphics{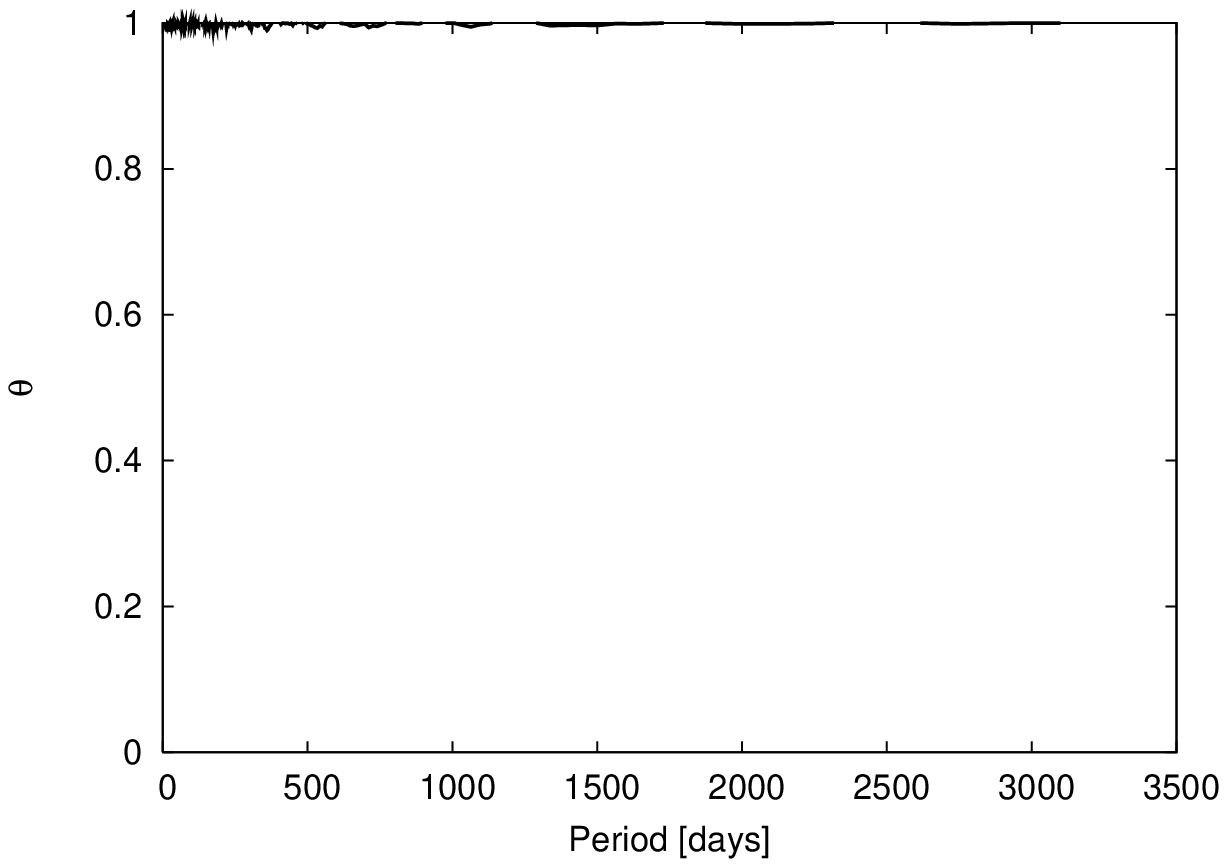}}\\
\caption{Periodograms determined using the Stellingwerf method. Parameter $\theta$ signifies the normalized phase variation. Upper left: Periodogram of mean equatorial zonal velocity. Upper right: Periodogram of mean equatorial meridional velocity. Bottom left: Periodogram of sampled heliographic latitude of the centre of the solar disc. Bottom right: Periodogram related to the sampling of data.}
\label{fig:periodograms}
\end{figure*}

We cannot completely exclude the influence of the calculated flow fields by the the position and orientation of the solar disc (position angle of the rotation axis $P$, heliographic latitude of the centre $b_0$), so we put also the series of both parameters sampled on the same dates when our measurements of surface flows exist through the period analysis. We also verified the periods caused by sampling of the data using the same method. Periodograms are displayed in Fig.~\ref{fig:periodograms}. 

We conclude that we did not detect any significant period in the available data set. As it can be seen in Fig.~\ref{fig:periodograms}, there exist non-convincing (the values of the parameter $\theta$ are quite high, which means that the period probably is not significant) signs of periods detected in the real data, which are not present in
the control data set. Values of the suspicious periods are 657~days (1.80~years) in the meridional component and 1712~days (4.69~years) in the zonal component. We note that the 1.8-year period was also detected by \cite{2004AA...418L..17K}. It is claimed to be related to a possible Rossby wave $r$-mode signature in the photosphere with azimuthal order $m \sim 50$ reported by \cite{2000Natur.405..544K}, but lately disputed e.\,g. by \cite{2006SPD....37.3002W}. The period estimate for such an $r$-mode is close to 1.8 years. According to \cite{2005AA...438.1067K}, such a periodicity was observed in the total magnetic flux only on southern hemisphere from 1997 to 2003. The coupling between the zonal flow and the meridional circulation could transfer the signal of the $r$-mode motion to the mean meridional component.

The detected suspicious periods may not be of solar origin. The sparse data set suffers from aliasing caused by a bad coverage of the studied interval. To confirm the periods far more homogenous data set is needed. This may be a task for ongoing space borne experiment Helioseismic Michelson Imager (HMI), which will be a succesor of MDI. 

Detected periodicities are absent in Mt.~Wilson torsional oscillations time series, which is far more homogenous material than the one used in this study. \cite{1990ApJ...351..309S} also did not find any time variations in the study tracking the features in the low resolution dopplergrams covering homogenously 20 years of Mt.~Wilson observations. All the arguments written above led us to leave the detected periods as suspicious, as they cannot be confirmed from the current data set.

\subsection{Relation to the magnetic activity}
We also investigated the coupling of equatorial zonal velocity (average equatorial solar rotation) and the solar activity in the near-equatorial area (belt of heliographic latitudes from $-10\,^\circ$ to $+10\,^\circ$). The average equatorial zonal velocity incorporates the average supergranular network rotation and also the movement of degenerated supergranules influenced by a local magnetic field with respect to their non-magnetic vicinity. Indexes of the solar activity were extracted from the daily reports made by \emph{Space Environment Center National Oceanic and Atmospheric Administration (SEC NOAA)}. Only the days when the measurements of horizontal flows exist were taken into account. As the index of the activity we have considered the total area of sunspots in the near-equatorial belt and also their type.

\begin{figure}
\resizebox{0.5\textwidth}{!}{\includegraphics{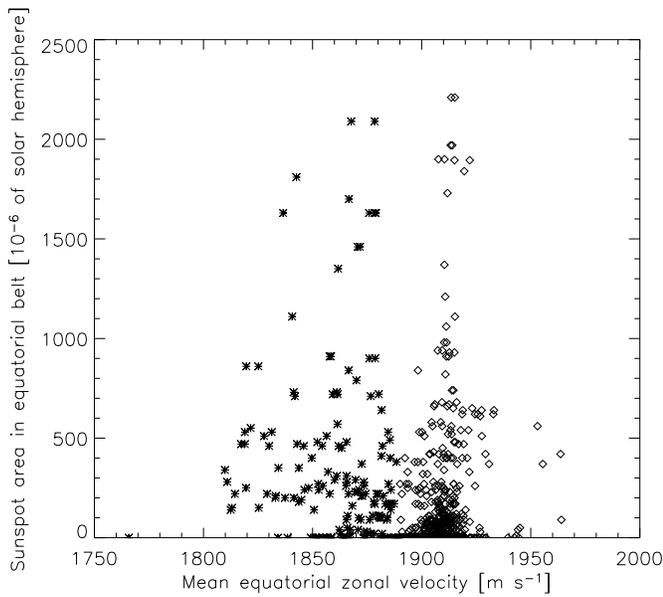}}
\caption{Mean zonal equatorial velocity versus the sunspot area in near-equatorial belt. We decided to divide the data in two regimes along the velocity axis. Although the division is arbitrary, we believe that it is supported by the theory of the dynamical disconnection of sunspots from their roots.}
\label{fig:activity_velocity}
\end{figure}

\begin{figure*}[!]
\resizebox{\textwidth}{!}{\includegraphics{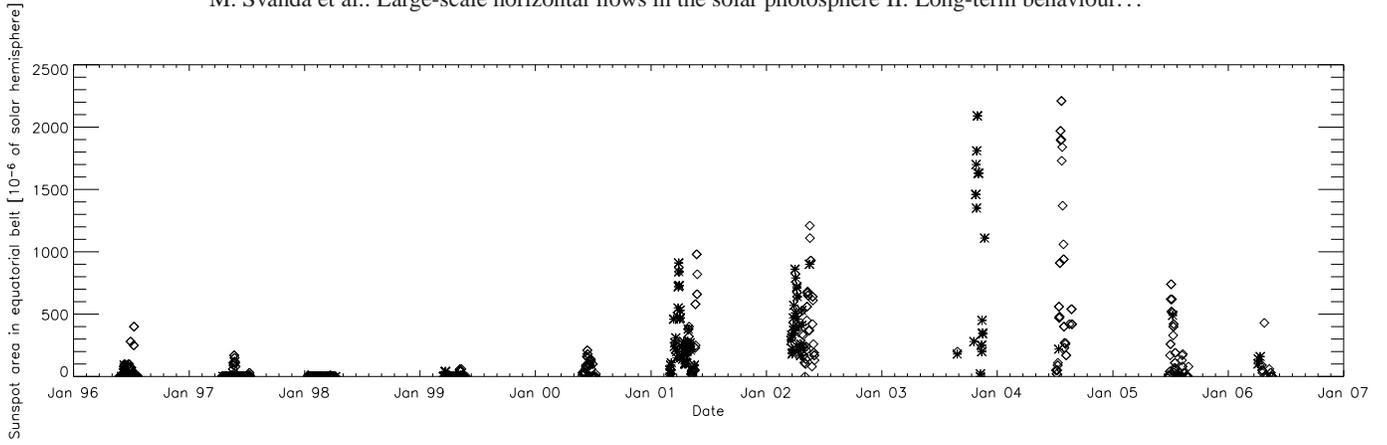}}
\caption{Sunspot area sampled in the same times when the measurements of the horizontal flows exist. Two regimes of the near-equatorial belt rotation are displayed. Diamonds denote ``fast'' rotating equatorial belts, crosses the ``scattered'' group.}
\label{fig:velocity_regimes}
\end{figure*}

First of all we computed the correlation coefficient $\rho$ between the mean equatorial zonal velocity and the sunspot area in the near-equatorial belt and got a value of $\rho=-0.17$. We cannot conclude that there is any linear relation between these two indices. The dependence of both quantities is plotted in Fig.~\ref{fig:activity_velocity}. We can clearly find two different regimes which are divided by the velocity of approximately 1890~\mps. In one regime (77~\% of the cases), the equatorial belt rotates about 60~\mps{} faster ($1910 \pm 9$~\mps; hereafter ``fast group'') than Carrington rotation, in the other one (23~\%) the rotation rate is scattered around the Carrington rate ($1860 \pm 20$~\mps; hereafter ``scattered group''). The division in two suggested groups using the speed criterion is arbitrary. If there exist only two groups, they certainly overlap and only a very detail study could resolve their members. We may also see more than two groups in Fig.~\ref{fig:activity_velocity}. The arguments for division in just two groups will follow.

For both regimes, there does not exist a typical sunspot area. The distribution of both regimes in time is displayed in Fig.~\ref{fig:velocity_regimes}. The data in the periods of larger solar activity (years 2001 and 2002, these are also the only years when the data cover two Carrington rotations continuously) show that both regimes alternate with a period of one Carrington rotation. 

The histogram of the mean zonal equatorial velocity has a similar, i.~e. bimodal, character like in Fig.~\ref{fig:activity_velocity} with a greater second peak, because such a histogram is constructed not only from belts cointaining magnetic activity but also from the belts where not magnetic activity was detected. The mean equatorial rotation for all the data is 1900~\mps, 1896~\mps{} for the equatorial rotation in a presence of sunspots, and 1904~\mps{} for days without sunspots in the equatorial region. 

Such bimodal velocity distribution is in disagreement with the results obtained by time-distance helioseismology by \cite{2004ApJ...607L.135Z}. In this work authors found that the stronger the magnetic field the faster such a magnetic element rotates. They observed about 70~\mps{} faster rotation than the average for magnetic areas with magnetic fields stronger than 600~G. It might be possible that the size of the magnetic area and its magnetic field strength play different roles in the influencing of the plasma motions. However, \cite{2005AA...436.1075M} showed that the size of the magnetic area and the maximum magnetic field strength or the total magnetic flux correlate quite well.

\begin{figure*}[!]
\resizebox{0.48\textwidth}{!}{\includegraphics{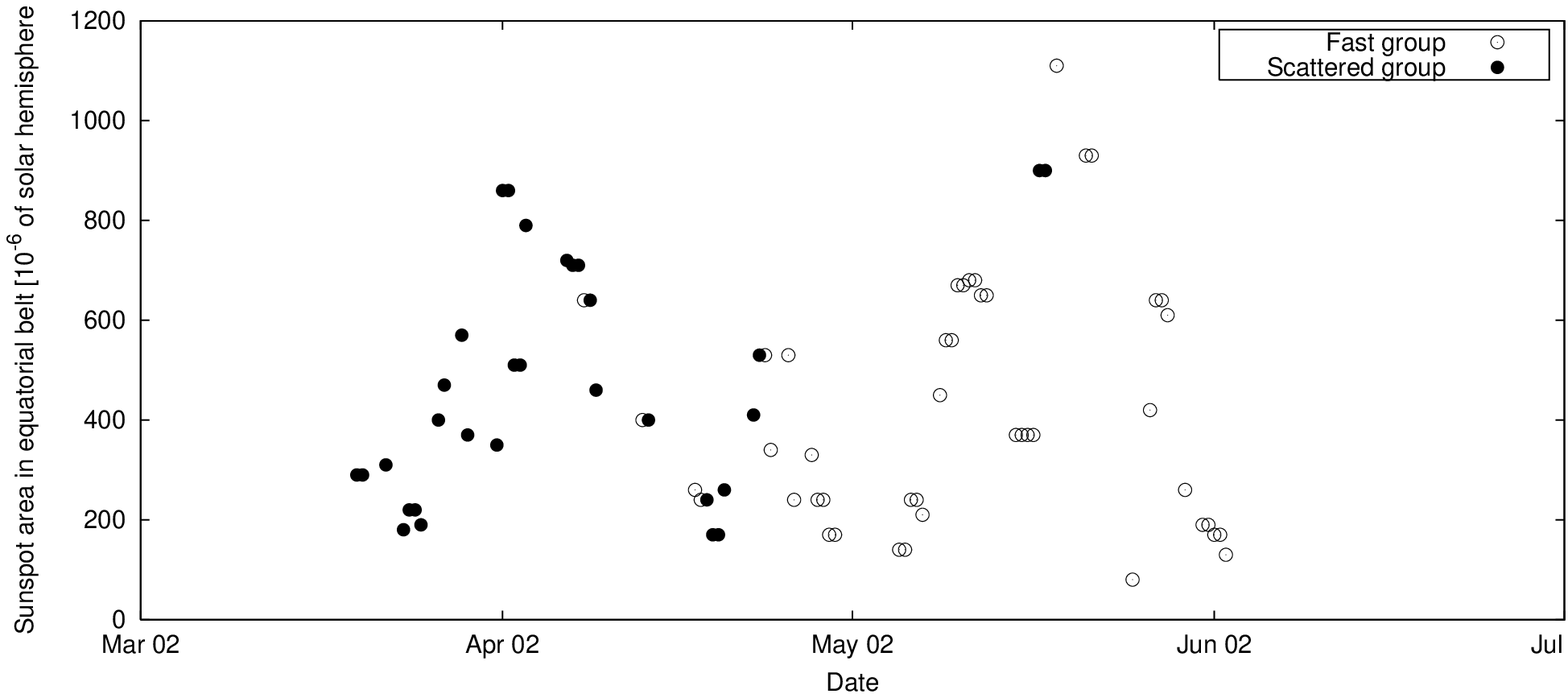}}
\resizebox{0.505\textwidth}{!}{\includegraphics{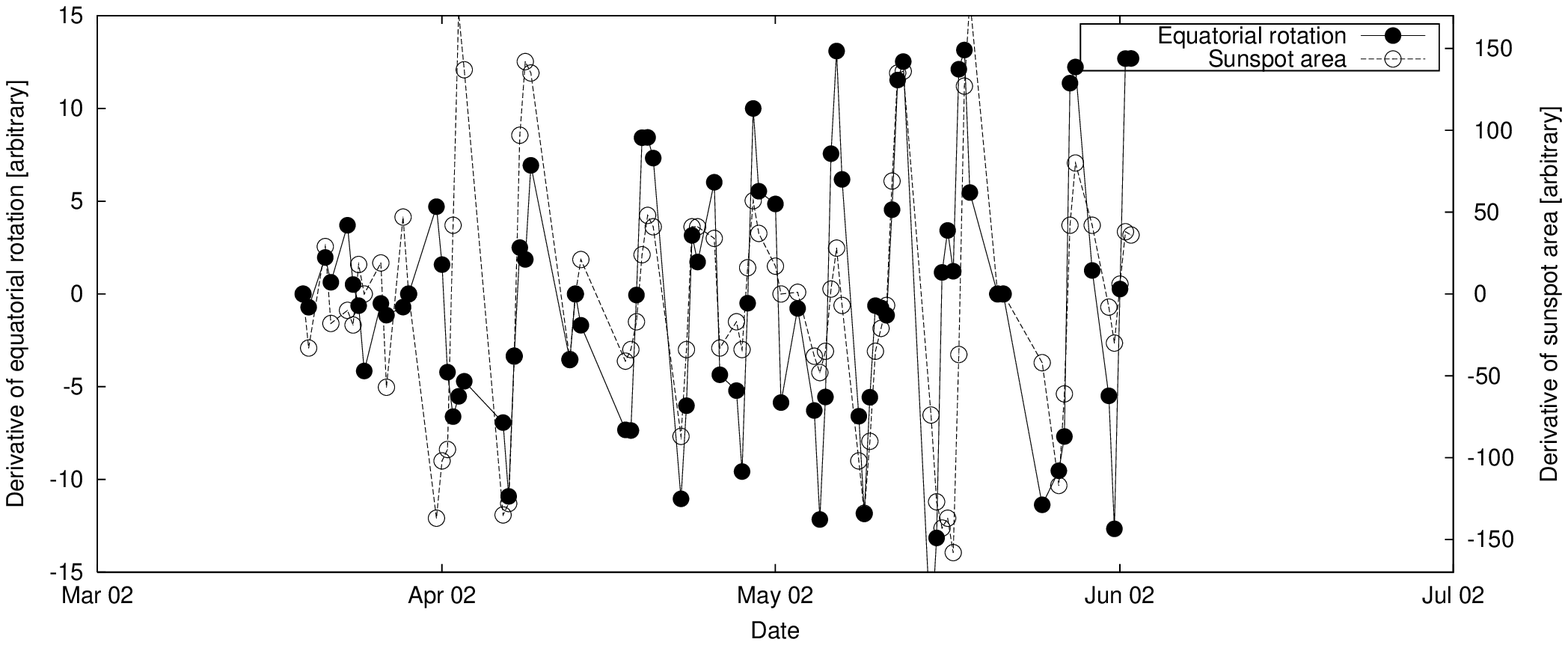}}
\caption{Left: Distribution of two equatorial rotation modes in year 2002. Right: Derivatives of the mean zonal velocity (solid curve) and the sunspot area in the near-equatorial region (dashed curve) in year 2002. Both quantities correlate with each other quite nicely.}
\label{fig:gradients2002}
\end{figure*}

\begin{figure*}[!]
\resizebox{0.48\textwidth}{!}{\includegraphics{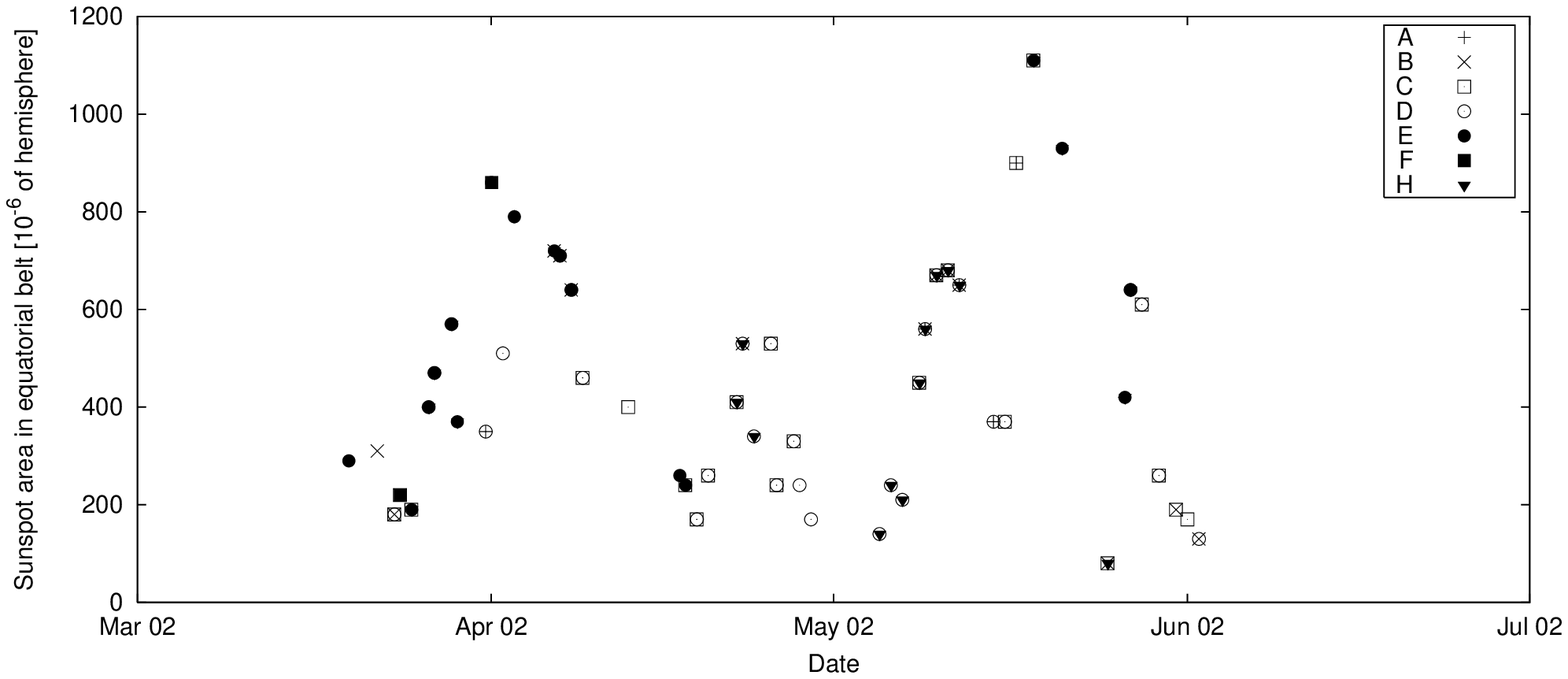}}
\resizebox{0.48\textwidth}{!}{\includegraphics{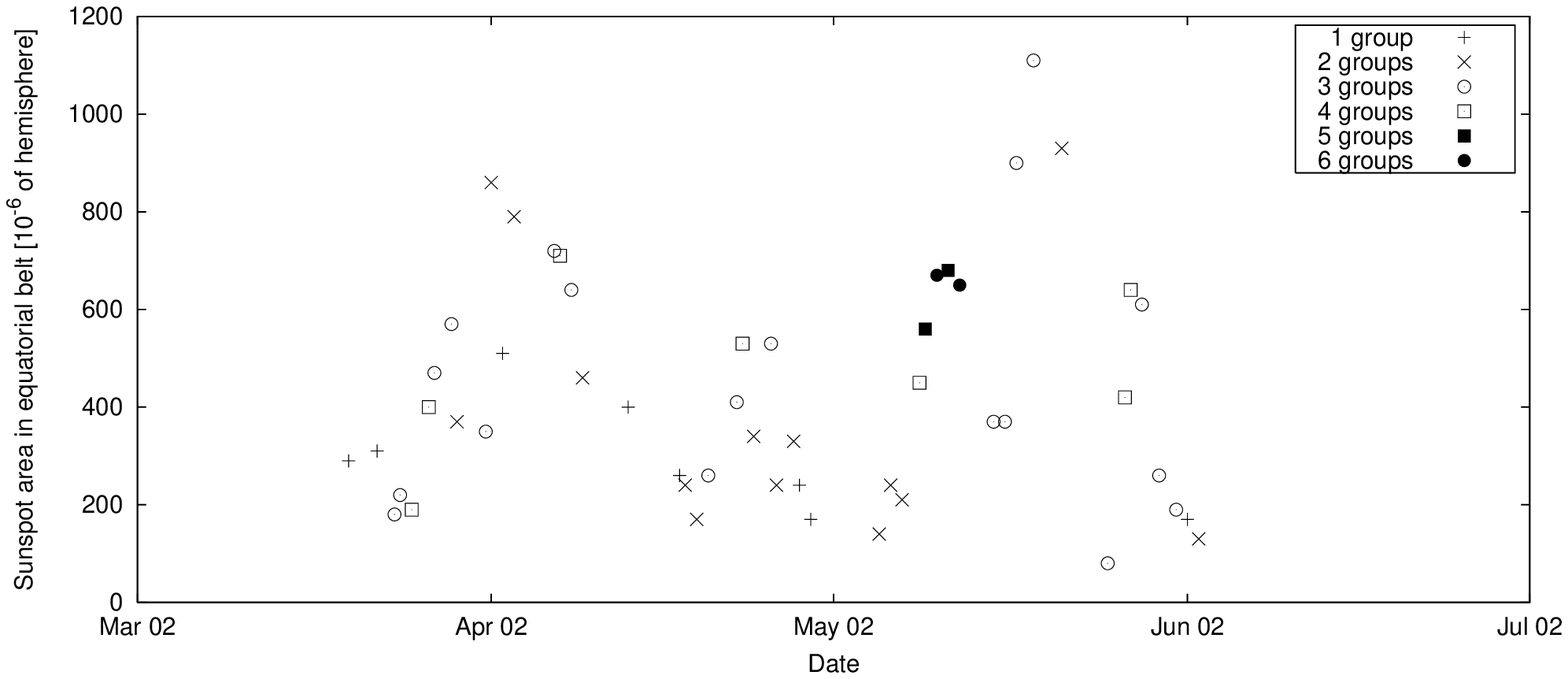}}
\caption{Active region morphological types (left) and the number of active regions in the near-equatorial belt distributed as a function of time and sunspot area.}
\label{fig:spottypes2002}
\end{figure*}

Detailed studies of the sunspot drawings obtained from the Patrol Service of Ond\v{r}ejov Observatory and the Mt.~Wilson Observatory drawings archive revealed that in the ``fast'' group, the new or growing young active regions were present in the equatorial belt. On the contrary, in the ``scattered'' group the decaying or recurrent active regions prevailed in the equatorial area. The deceleration of the sunspot group with its evolution was noticed e.\,g. by \cite{2004SoPh..221..225R}. Moreover, our results suggest that the new and rapidly growing sunspots in the studied sample (March to May 2001 and April to June 2002) move with the same velocity. This behaviour could be explained by emergence of the local magnetic field from a confined subphotospheric layer. According to the rough estimate \citep{2002SoPh..205..211C} the speed of $1910 \pm 9$~\mps{} corresponds to the layer of $0.946 \pm 0.008\ R_\odot$, where the angular velocity of rotation suddenly changes. During the evolution, the magnetic field is disrupted by the convective motions. An interesting behaviour is displayed by the alternation of the ``fast'' and ``scattered'' regimes (see Fig.~\ref{fig:gradients2002}) with the period of one Carrington rotation. It suggests that active regions in the equatorial region emerge in groups. We have to keep in mind that our study produces a very rough information due to the averaging of all efects in the equatorial belt. 

The observed behaviour could be a manifestation of the disconnection of magnetic field lines from the base of the surface shear during evolution of the growing sunspot group. This behaviour was theoretically studied by \cite{2005AA...441..337S}. They suggested the dynamical disconnection of bipolar sunspot groups from their magnetic roots deep in the convection zone by upflow motions within three days after the emergence of the new sunspot group. The motion of sunspots changes during those three days from ``active'' to ``pasive''. The active mode is displayed by motions reasonable faster with respect to a non-magnetic origin. The passive mode means mostly the deceleration of sunspot motions and influence of the sunspot motions only by the shallow surface plasma dynamics. The theory of the disconnection of sunspot groups from their magnetic root supports the division of the data set in two groups. 

\begin{figure}
\resizebox{0.49\textwidth}{!}{\includegraphics{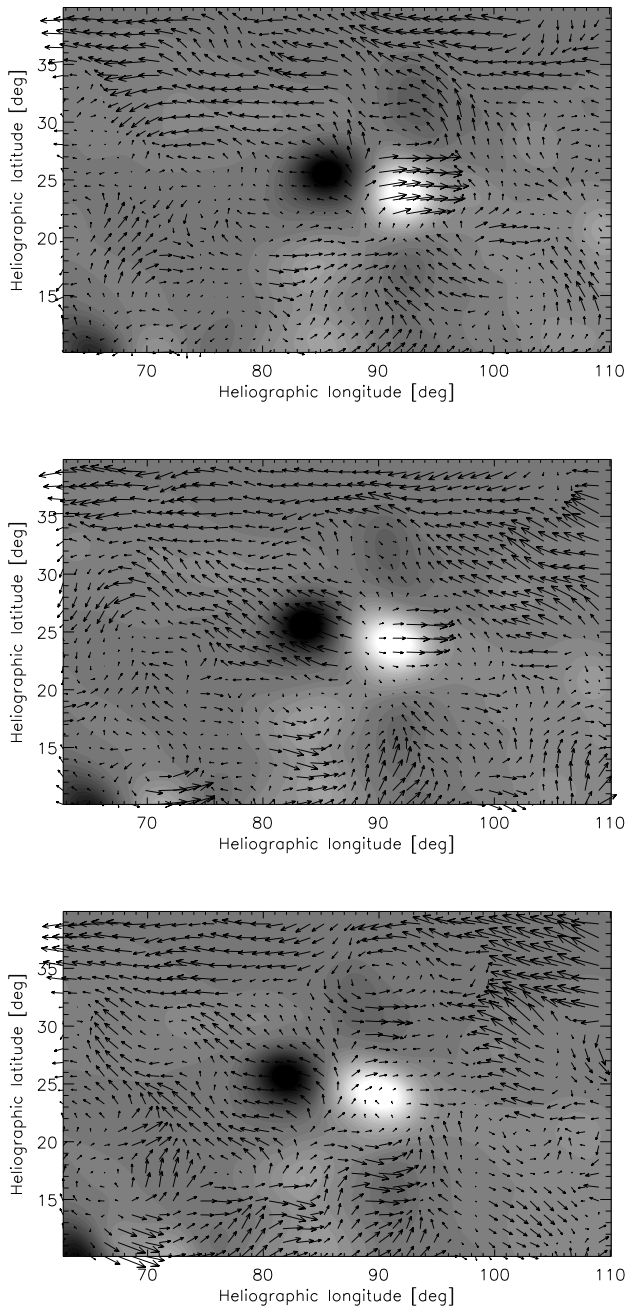}}
\caption{Case study: Evolution of the flows in and around the active region NOAA~9368 on March 6, March 7, and March 8, 2001. The leading polarity rotate significantly faster than the following one and the non-magnetic surroundings in the first day. The whole group slows down in the other two days. As the background image, the MDI magnetogram smoothed to the resolution of the measured flow field is used. The magnetic field intensities are displayed in the range of $-$1800 (black) to $+$1800 (white) Gauss in the linear scale.}
\label{fig:group_evolution}
\end{figure}

As an example, we selected the active region NOAA~9368 (Fig.~\ref{fig:group_evolution}) to show the behaviour of large-scale velocities in time. We see that the leading part of the active region rotates faster than the surroundings in the first day of observation and the whole group slows down in next two days. The inspection of details in the behaviour of selected active regions in the whole data set will be the subject of the ongoing studies.

Division of the equatorial belt into 10 sectors helped the investigation of the behaviour of different active region types. The mean zonal velocity in the sector containing the studied active region according to morphological type of active regions are summarized in Table~\ref{tab:sunspot_types}. It can be clearly seen that, in average, more evolved active regions rotate slower than less evolved or young active regions, what is in agreement with e.~g. \cite{1986AA...155...87B}. We have also replotted Fig.~\ref{fig:activity_velocity} using segmented equatorial belt and different active region types and got a very similar result. We did not find any particular behaviour for various active region types (see Fig.~\ref{fig:spottypes2002}).
\begin{table}[b]
\caption{Mean synodic rotation velocities of different active regions types and the average equatorial rotation of all data. The measured speed values may be systematically biased by the projection effect \cite[see][]{2006ApJ...644..598H}.}
\centering
\begin{tabular}{cc}
\hline
\hline
Sunspot & Mean rotation \\
type & [\mps]\\
\hline
A & $1893 \pm 48$ \\
B & $1893 \pm 49$ \\
C & $1890 \pm 71$ \\
D & $1880 \pm 73$ \\
E & $1880 \pm 50$ \\
F & $1874 \pm 73$ \\
H & $1872 \pm 51$ \\
\hline
Average & $1900$ \\
\hline
\end{tabular}
\label{tab:sunspot_types}
\end{table}

We have also focused how the presence of the magnetic active areas will influence the average flow field. Since we found that a direct correlation is weak due to existence of two different regimes, we decided to study the temporal change of both quantities. The aim is to study whether an emerging active region in the near-equatorial belt will influence the average equatorial rotation. We computed numerical derivatives of the total sunspot area in the near-equatorial belt and of the average zonal equatorial flow. We have found that the correlation coefficient between both data series is $\rho=0.36$ and is higher for the ``fast group'' ($\rho=0.41$) than for the ``scattered group'' ($\rho=0.24$). The correlation is higher in periods of increased magnetic activity in the equatorial belt. For example, for data in year 2001 the correlation coefficient is $\rho_{2001}=0.58$ and for year 2002 $\rho_{2002}=0.52$; see Fig.~\ref{fig:gradients2002}. In both particular cases, the correlation is higher for the ``fast regime'' ($\rho \sim 0.7$) than for the second group. 

It is important to know that LCT in our method measures basically the motions of supergranules influenced by the magnetic field, not by spots that are recorded in the used solar activity index. Therefore the results can be biased by the fact that the presence of magnetic field does not necessarily mean the presence of sunspot in the photosphere. 

We think that despite an apparent disagreement, our apparently conflicting results can be valid and are in agreement with the results published earlier. Basically, as described by e.\,g. \cite{1990ApJ...357..271H} and explained in the model of \cite{2004SoPh..220..333B}, the solar rotation in the lower latitudes is slower in the presence of magnetic field. This is summarized in Table~\ref{tab:sunspot_types} and displayed in Fig.~\ref{fig:activity_velocity}. In most cases, ``spotty'' equatorial belts seem to rotate slower than the average for the whole data series. However, it is clear that emerging active regions cause in most cases the increase of the rotation rate. This is in agreement with a generally accepted statement found first by \cite{1970SoPh...12...23H} and \cite{1978ApJ...219L..55G}. The relation, obtained using linear fit on our data set, can be described by the equation
\begin{equation}
\Delta v \sim 0.2\, \Delta A_{\rm sunspots}\ {\rm m\,s^{-1}},
\end{equation}
where $\Delta v$ is a change of the equatorial rotation speed with respect to the Carrington rotation and $\Delta A_{\rm sunspots}$ is a relative change of sunspot area in the equatorial belt (in 10$^{-6}$ of solar hemisphere). We estimate that strong local magnetic areas rotate few tens of \mps{} faster than the non-magnetic surroundings.

The difference in the behaviour of the flow fields in the regions occupied by the magnetic field and their vicinity was studied also e.\,g. by  \cite{1992ApJ...393..782T} using the high-resolution data obtained at the Swedish Vacuum Solar Telescope on La Palma, Canary Islands. The authors found that the magnitude of horizontal velocities measured by LCT on the granular scale is in the regions of the quiet Sun larger than in an active-region plage. The high-resolution velocity fields are of different nature than the large-scale ones studied by our method. Flow fields on granular scales are mostly chaotic due to the turbulent behaviour. In the regions occupied by the magnetic field, the motions become more organized, the chaotic component is suppressed, and therefore the amplitude of the horizontal velocity is generally lower.
 
\cite{1990ApJ...351..309S} tracked the features in the low-resolution dopplergrams measured at the Mt.~Wilson 46~m tower telescope for the period of 20~years. They interpreted the detected flows as the velocity of the supergranular network, although the spatial resolution was lower than the size of individual supergranules. They found two regimes taking place in the rotation of the photosphere -- the quiet Sun and active regions. The results showed that the regions occupied by the magnetic field display a slower rotation than the non-magnetic vicinity. In general, the results of this study are in agreement with the results of the current one. \citeauthor{1990ApJ...351..309S} found the mean rotational rate of magnetic regions with value of (1864$\pm$1)~\mps, which, within the statistical errors, agrees with our findings. However, the synodical equatorial rotation of all Doppler structures measured by \citeauthor{1990ApJ...351..309S} is (1924$\pm$6)~\mps, which is more than 1~\% faster than the average rate measured in the present study. We could probably explain such a disagreement by the different resolution of the used data. In the study of \cite{1990ApJ...351..309S}, the full solar disc was sampled in a 34$\times$34 pixels array, while in the current one, the size of the solar disc was 1000$\times$1000 pixels. We assume that this faster rotational rate is due to the effect of undersampling of Doppler structures in quiet regions. The results of our study should not be influenced by this effect, since the supergranular structures are well resolved in the data.

\section{Conclusions}
We have verified that the method developed and tested using the synthetic data (Paper~I) is suitable for application to real data obtained by the MDI on-board SoHO and maybe also to the data that will be produced by its successor Helioseismic Michelson Imager (HMI) on-board the Solar Dynamic Observatory (SDO). HMI will have a greater resolution and will cover larger time span than two months each year. We verified that the long-term evolution of the horizontal velocity fields measured using our method is in agreement with generally accepted properties. 

During the periodic analysis of the equatorial area we found two suspicious periods in the real data, which are not present in the control data set containing the inclination of the solar axis towards the observer, the quantity that can bias systematically and periodically the results by a few \mps. The periods of 1.8~year and 4.7~years need to be confirmed using a more homogenous data set.

We also found that the presence of the local magnetic field generally speeds-up the region occupied by the magnetic field. However, we cannot conclude that there exists a dependence of this behaviour for different types of sunspots. We can generally say that the more evolved types of active regions rotate slower than the young ones, however the variance of the typical rotation rate is much larger than the differences between the rates for each type. We have found that the distribution of active regions rotation is bimodal. The faster-rotating cases correspond to new and growing active regions. Their almost constant rotation speed suggests that they emerge from the base of the surface radial shear at $0.95\ R_\odot$. The decaying and recurrent regions rotate slower with a wider scatter in their velocities. This behaviour suggests that during the sunspot evolution, sunspots loose the connection to their magnetic roots. Both regimes alternate with a period of approximately one Carrington rotation in years 2001 and 2002, which suggests that new active regions emerge in groups and may have a linked evolution.

\begin{acknowledgements}
The authors of this paper were supported by the Czech Science Foundation under grants 205/03/H144 (M.~\v{S}.) and 205/04/2129 (M.~K.), by the Grant Agency of the Academy of Sciences of the Czech Republic under grant IAA 3003404 (M.~S.) and by ESA-PECS under grant No. 8030 (M.~\v{S}.). The Astronomical Institute of Academy of Sciences is working on the Research project AV0Z10030501 of the Academy  of Sciences, Astronomical Institute of Charles University on the Research Program MSM0021620860 of the Ministry of Education. The MDI data were kindly provided by the SoHO/MDI consortium. SoHO is the project of international cooperation between ESA and NASA. Authors would like to acknowledge the staff of W.~W.~Hansen Experimental Physics Laboratory, computer resources of which were used during the data processing. We thank to the referee Roger~K.~Ulrich, whose useful comments significantly improved the quality of the paper.
\end{acknowledgements}

\newcommand{\SortNoop}[1]{}

\end{document}